\documentclass[sigconf]{acmart}
\usepackage{booktabs}
\usepackage{enumitem}
\usepackage{float}
\usepackage{flushend}
\usepackage{graphicx}
\usepackage{hyperref}
\usepackage{makecell}
\usepackage{multirow}
\usepackage{subcaption}
\usepackage[most]{tcolorbox}
\usepackage{xcolor}

\AtBeginDocument{%
  }

\copyrightyear{2025}
\acmYear{2025}
\setcopyright{acmlicensed}\acmConference[UIST '25]{The 38th Annual ACM Symposium on User Interface Software and Technology}{September 28-October 1, 2025}{Busan, Republic of Korea}
\acmBooktitle{The 38th Annual ACM Symposium on User Interface Software and Technology (UIST '25), September 28-October 1, 2025, Busan, Republic of Korea}
\acmDOI{10.1145/3746059.3747689}
\acmISBN{979-8-4007-2037-6/2025/09}

\acmSubmissionID{3953}

\begin{document}

\title{ViseGPT: Towards Better Alignment of LLM-generated Data Wrangling Scripts and User Prompts}

\author{Jiajun Zhu}
\authornote{Both authors contributed equally to this research.}
\email{jiajunzhuchris@zju.edu.cn}
\orcid{0009-0002-1438-0561}
\author{Xinyu Cheng}
\authornotemark[1]
\email{pikabi@zju.edu.cn}
\orcid{0009-0002-1161-1301}
\affiliation{%
  \institution{State Key Lab of CAD\&CG, \\ Zhejiang University}
  \city{Hangzhou}
  \state{Zhejiang}
  \country{China}
}

\author{Zhongsu Luo}
\orcid{0009-0003-0885-2742}
\affiliation{%
  \institution{State Key Lab of CAD\&CG, \\ Zhejiang University}
  \city{Hangzhou}
  \state{Zhejiang}
  \country{China}
}
\email{zhongsuluo@zju.edu.cn}

\author{Yunfan Zhou}
\orcid{0009-0009-7814-4390}
\affiliation{%
  \institution{State Key Lab of CAD\&CG, \\ Zhejiang University}
  \city{Hangzhou}
  \state{Zhejiang}
  \country{China}
}
\email{yunfzhou@zju.edu.cn}

\author{Xinhuan Shu}
\orcid{0000-0002-9736-4454}
\authornote{Di Weng and Xinhuan Shu are the co-corresponding authors.}
\affiliation{%
  \institution{Newcastle University}
  \city{Newcastle Upon Tyne}
  \country{United Kingdom}}
\email{xinhuan.shu@newcastle.ac.uk}

\author{Di Weng}
\orcid{0000-0003-2712-7274}
\authornotemark[2]
\affiliation{%
  \institution{School of Software Technology, \\ Zhejiang University}
  \city{Ningbo}
  \state{Zhejiang}
  \country{China}
}
\email{dweng@zju.edu.cn}

\author{Yingcai Wu}
\orcid{0000-0002-1119-3237}
\affiliation{%
  \institution{State Key Lab of CAD\&CG, \\ Zhejiang University}
  \city{Hangzhou}
  \state{Zhejiang}
  \country{China}
}
\email{ycwu@zju.edu.cn}

\renewcommand{\shortauthors}{Zhu et al.}
\newcommand{\lzs}[1]{\textcolor{orange}{[lzs: #1]}}

\begin{abstract}
% Large language models (LLMs) enable the rapid generation of data-wrangling scripts through natural language interaction. 
% However, these scripts often contain implicit issues that prevent them from fully adhering to user-specified requirements, necessitating careful inspection and iterative refinement.
Large language models (LLMs) enable the rapid generation of data-wrangling scripts based on natural language instructions, but these scripts may not fully adhere to user-specified requirements, necessitating careful inspection and iterative refinement.
% Existing approaches typically require users to manually analyze script logic rather than providing direct validation of correctness.
Existing approaches primarily assist users in understanding script logic and spotting potential issues themselves, rather than providing direct validation of correctness.
To enhance debugging efficiency and optimize the user experience, we develop ViseGPT, a tool that automatically extracts constraints from user prompts to generate comprehensive test cases for verifying script reliability. 
% The test results are then transformed into an interactive visual representation, allowing users to intuitively assess alignment with semantic requirements and iteratively refine their scripts. 
The test results are then transformed into a tailored Gantt chart, allowing users to intuitively assess alignment with semantic requirements and iteratively refine their scripts. 
Our design decisions are informed by a formative study (N=8) that explores user practices and challenges. 
% We further developed a prototype system, ViseGPT, and evaluated its effectiveness and usability through a user study (N=18). 
We further evaluate the effectiveness and usability of ViseGPT through a user study (N=18). 
Results indicate that ViseGPT significantly improves debugging efficiency for LLM-generated data-wrangling scripts, enhances users' ability to detect and correct issues, and streamlines the workflow experience.
\end{abstract}




\begin{teaserfigure}
    \centering
    \includegraphics[width=1\columnwidth]{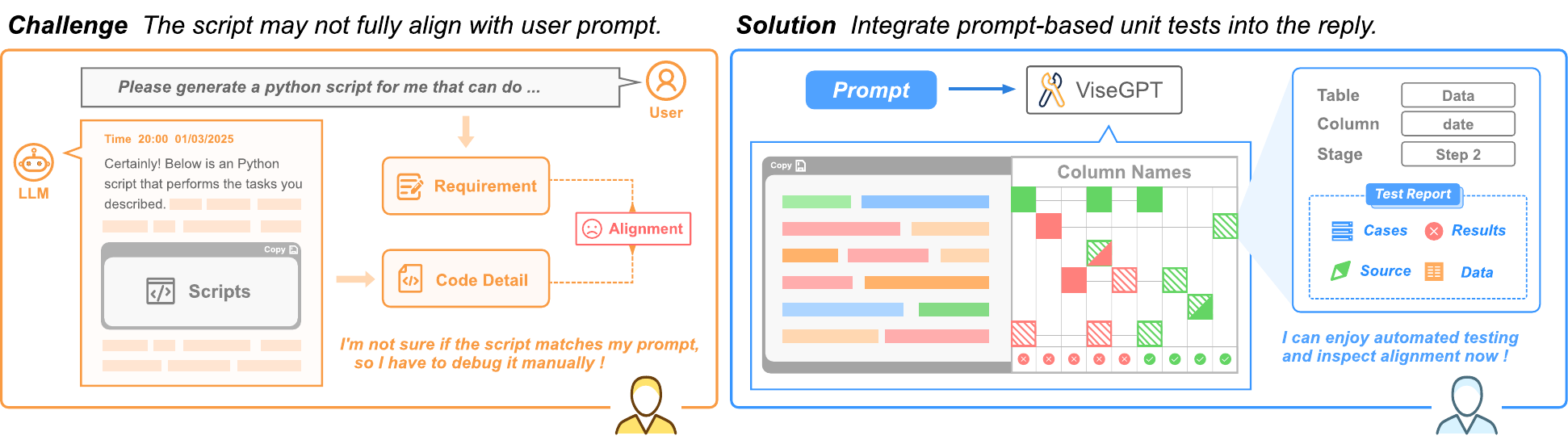}
    \caption{ViseGPT automatically extracts test cases from the user's prompts for efficient and accuracy debugging.}
    \label{fig:teaser}
\end{teaserfigure}

% \centering
%   \includegraphics[width=1.9\columnwidth]{figures/model_workflow.pdf}
%   \caption{Workflow of the model}\Description{This figure illustrates the workflow of the model, detailing the steps and processes involved.}


\maketitle

\section{Introduction}

Data wrangling is a process of making raw data suitable for use and analysis~\cite{SeanKandel2011wrangling}.
Due to the complexity of data wrangling processes, one common approach is to write custom scripts to perform data cleaning and transformation operations.
With recent advances in large language models (LLMs), these data wrangling scripts can be easily generated based on the given instructions, substantially lowering
% many models have been developed to assist users in generating data wrangling scripts based on the given instructions, reducing the need for manual coding, which substantially lowers 
the barrier for less-experienced users to complete complex data wrangling tasks~\cite{10.1145/3650203.3663334,10555056,10.1145/3691620.3695267,10.14778/3685800.3685890}.
However, LLM-generated scripts often raise reliability concerns, such as latent defects \cite{mcaleese2024llmcriticshelpcatch,chen2024witheredleaffindingentityinconsistencybugs}, hallucinations \cite{10.1145/3703155,ji-etal-2023-towards}, ambiguities \cite{10825265,mehrparvar-pezzelle-2024-detecting}, and mismatches with user requirements~\cite{ma2024engineerpromptstraininghumans,jiang2025chatbugcommonvulnerabilityaligned}, which demand effective approaches to assist users in reviewing and validating these scripts before use. 

To help users understand and validate data wrangling scripts, some efforts have been made, such as displaying the script transformation pipeline \cite{9693232} and directly assisting users in comprehending the input and output space of the scripts \cite{10670464}.
% And towards better understanding of LLM-generated scripts, recent works have introduced tools
Recent works improve understanding of LLM-generated scripts by introducing tools that represent the inherent data transformation process, including visualizing the behavior of real-time generated scripts as a tree diagram ~\cite{Xie_2024} and supporting nodes querying for the operation details \cite{10026499}. 
However, users are still required to manually validate whether the LLM-generated scripts can execute correctly and fulfill task-specific requirements.
Our formative study further reveals that users often do not seek a comprehensive understanding of the script itself, but rather a direct way to evaluate whether the script aligns with their requirements.

In this paper, we aim to improve the alignment between users' instructional prompts and LLM-generated data wrangling scripts and facilitate efficient verification and maintenance of the functional correctness of these scripts.
% and addressing identified issues, with the overarching goal of enhancing alignment between scripts and user prompts. 
% Our goal is to allow users to quickly verify the reliability, identify issues, and correct discrepancies in the LLM-generated data wrangling scripts by comparing the generated scripts with the information involved in the prompt.
% However, challenges arise when we attempt to both obtain the test cases and enhance the user's work efficiency. 
% To improve efficiency, practitioners often opt to sample only a portion of the data before obvious issues are identified, spending significant time tracing back only when problems arise. 
% Hence, we employ unit testing methods (e.g., \cite{10.1145/3691620.3695501,10329992}) to validate that each step produces the expected results according to the predefined constraints extracted from the user's natural language requirements. 
Inspired by the unit testing approach~\cite{10.1145/3691620.3695501,10329992},
we propose ViseGPT, a tool that uses automatically generated test cases based on given prompts to assist users in validating LLM-generated data wrangling scripts.
% we propose ViseGPT, a tool that assists users in validating LLM-generated data wrangling scripts with test cases automatically generated based on the given prompts. 
% ), a common method in software engineering that validates program correctness by generating test cases and executing systematic tests automatically. 

% Its approach of atomizing program steps and chaining them together is helpful for locating issues, improving efficiency and lowering the barrier to entry.
% In software development, common unit testing practices often utilize assertion-based validation through actual value comparison. 
% However, challenges arise if we attempt to obtain concrete test data and assertions as the test cases for unit testing. 
However, leveraging the unit testing approach for LLM-generated data wrangling scripts presents three key challenges: 
\begin{itemize}[leftmargin=10pt]
\item First, generating test cases that accurately reflect user requirements from natural language prompts is non-trivial. 
User prompts often contain implicit information and domain-specific knowledge that are difficult to extract. 
For example, a prompt like ``remove outliers from the sales data'' may imply statistical rules (e.g., excluding negative values) that are not explicitly stated. 
Traditional unit testing relies on precise specifications, but natural language prompts are inherently ambiguous, making automated test case generation prone to misinterpretation. 
\item Second, validating whether the generated script fully complies with the prompt requires more than simple input-output comparisons. 
Unlike conventional programming where expected outputs can be predefined, data wrangling tasks often involve complex transformations where intermediate steps influence correctness. 
% For instance, a script that merges two datasets must not only produce the correct final output but also preserve key relationships between columns. 
Existing testing frameworks focus on functional correctness at the ends but lack mechanisms to assess higher-level semantic alignment with user intent. 
\item Third, guiding users in refining scripts based on test failures demands an interpretable feedback mechanism. 
When a script fails to meet certain conditions, users need actionable insights to adjust the script. 
However, with limited contextual information, users have to manually trace issues according to data flow. 
This becomes particularly challenging for less-experienced users who may struggle to map test failures back to the original requirements or identify missing edge cases. 
Without structured guidance, iterative improvement devolves into a trial-and-error process, reducing efficiency gains.
\end{itemize}
To address these challenges, we developed ViseGPT, a tool that generates test cases based on user prompts, presenting test results within an interactive visualization interface centered around a tailored Gantt chart \cite{WILSON2003430} and supporting iterative sending of expanded prompts for ideal results.
Specifically, the system incorporates an analysis model that can extract constraints about LLM-generated data wrangling scripts \cite{10670464} from users' natural language prompts and process these into systematic test cases, which are then automatically run against the LLM-generated scripts to validate their correctness. 
This constraint-based approach allows ViseGPT to evaluate whether generated scripts semantically align with user requirements. 
The test results are presented through an intuitive visualization interface featuring a tailored Gantt chart that leverages symbolic representations of column data transformations and color-coded test status indicators, enabling users to quickly grasp script behavior and identify issues. 
Furthermore, the interactive interface supports deeper exploration, allowing users to examine detail data at each processing step, review associated test cases and make custom adjustments to existing constraints. 
To integrate with the test framework, ViseGPT allows users to refine scripts by feeding test reports back into the LLM as expanded prompts for improved script regeneration. 
This closed-loop system not only helps users verify functional correctness of scripts but also provides actionable feedback for continuous programming improvement.
We further evaluate ViseGPT through a user study (N=18), showing improvement and inspiration in debugging LLM-generated data wrangling scripts efficiently and accurately.
% In summary, our contributions are three-fold.
% \begin{itemize}[leftmargin=20pt]
% % Theory & Experiment
% % \item A theoretical approach that enables direct unit testing based on constraint extracted from natural language, thereby improving the efficiency of user programming process. % Discussion and experimental verification may be required
% % Model
% \item We propose a constraint-based model that extracts test cases from user prompts to inspect data wrangling scripts.
% % System
% \item We develop ViseGPT, a tool that assists users in efficiently debugging LLM-generated data wrangling scripts.
% % Evaluation
% \item We evaluate ViseGPT through an user study (N=18), showing improvements and inspirations in debugging LLM-generated data wrangling scripts efficiently and accurately.
% \end{itemize}
\section{Related Work}

We review LLM-assisted data wrangling script generation, unit test validation, and human-LLM interactive interface design, which are closely related to our study.

\subsection{LLM-generated Data Wrangling Scripts}
% Overview of how LLMs have transformed data wrangling by lowering technical barriers
Data wrangling often involves dealing with large volumes of intricate data and varying scripts tailored to different scenarios \cite{Wrex, Unravel, Mayer2015User}. 
Traditionally, the approach requires users to have programming skills to complete the task. 
Within the data analysis workflow, this is a tedious and error-prone task that can consume up to 80\% of a worker's time and effort \cite{10.1145/1978942.1979444, Dasu2003, 10670464}. 
% Facing these issues, there is enthusiasm for using LLM agent during data wrangling script programming to reduce the learning barrier and enhance work efficiency
Facing these issues, researchers are enthusiastic about using an LLM agent in data wrangling script programming to reduce the learning barrier and improve efficiency \cite{Anthropic2025, Google2024, OpenAI2024}. 
In the HCI community, there is a long history of assisting users in understanding, validating, and refining data wrangling scripts \cite{history1, history2, history3, history4, history5}.

% They need to be optimized on the basis of the existing tools, and there are three aspects of improvement
However, it is impractical to expect these native LLM agents to directly produce correct and user-demand-satisfying data wrangling scripts due to the inherent ambiguities in natural language \cite{10825265,mehrparvar-pezzelle-2024-detecting} and the mechanisms of the large models themselves \cite{10.1145/3703155,ji-etal-2023-towards,ma2024engineerpromptstraininghumans,jiang2025chatbugcommonvulnerabilityaligned}. 
By taking advantage of the LLM agent itself, some works can effectively help users obtain more accurate data wrangling scripts \cite{SheetAgent, Sheetcopilot}.
In conclusion, current improvement strategies focus on three aspects that display the data transformation process, customize prompts, and backtest generated results.

\textbf{Displaying the process of data transformation. }In the context of data wrangling, it is common to use a script to progressively process one or more tabular data objects \cite{Dasu2003}. Displaying the detailed changes in the data at each step can help users understand the script's logic and assist in debugging. 
For example, in XNLI by Feng et al. \cite{10026499}, users can query nodes during the data transformation process. 
In WaitGPT \cite{Xie_2024}, the script is dynamically updated in a tree structure, providing real-time visualization of the transformations. 
And in Wrangler \cite{10.1145/1978942.1979444}, the system allows users to directly manipulate the individual steps in the data transformation process. 
Through reasoning and inspection, users can obtain the logical chain of the script, but this approach remains insufficiently intuitive to present the functional correctness of scripts. 
When dealing with more complex scripts, users need significant analytical skills to identify issues within the script.

\textbf{Customizing prompts.} For LLMs, specific and well-structured prompts can reduce the degree of freedom in generation, thus improving the accuracy of the results \cite{liu2022structuredprompttuning}. % Customized prompts allow users to submit prompts in a particular format or provide more detailed background information, helping LLMs generate result that better meet their requirements.
In Dango \cite{Dango}, users clarify their intent by answering multiple-choice questions posed by the LLM, and receive multiple forms of feedback to aid evaluation, thereby enhancing user communication. 
In the work by Liu et al. \cite{10.1145/3544548.3580817}, users can opt to break down tasks and input prompts step by step when the LLM-generated results are not satisfactory. 
And in ColDeco \cite{10305647}, users can decompose a generated solution into intermediate helper columns to understand how the problem is solved step by step. 
Furthermore, in DIY \cite{10.1145/3397481.3450667}, a sandbox is provided for users to interact with information about queries and a subset of the database, in which end-users can evaluate the system’s response. 
Clearly, rich and structured prompts can lead LLMs to generate more demand-satisfying data wrangling scripts. 
However, this approach increases the cooperating burden on users to preprocess and structure their prompts.% Additionally, methods involving task decomposition or supplementary information are not always feasible \cite{10.1145/3544548.3580817}.

\textbf{Backtesting the generated results.} To validate script correctness, data test sets can be generated based on the prompt or the predefined functionality of the code. 
Upon obtaining the test results, the report can be provided to the user, or automated iterative modifications can be performed until the accuracy reaches a predefined threshold. 
In SpoC \cite{NEURIPS2019_7298332f}, the system predicts the program line responsible for the failure and focuses the search on alternative translations of that pseudocode line. 
Additionally, CodeScore \cite{dong2024codescoreevaluatingcodegeneration} implements an approach to assess code generation by predicting code execution. 
The quality of the dataset is fundamental to traditional unit testing methods.
% And if the scale of test cases that are too small may fail to cover edge cases, while overly large-scale validation will require excessive time in repeated script executions
If the test case scale is too small, it may fail to cover edge cases, while an excessively large validation set will require substantial time for repeated script executions \cite{10.1145/2685612}. 
But this approach can be used to align scripts with user requirements and provides valuable insights. 

\subsection{Unit Testing for Script Validation}
Unit testing refers to the process of testing individual hardware or software units or groups of related units \cite{159342}. 
In practice, developers insert assertions \cite{10795010,10.1145/3377811.3380429,alagarsamy2024a3test} or logical statements within code segments \cite{7163725,xie2006tool} to verify whether the program's execution matches their expectations. 
These assertions are crucial for identifying issues early in the development phase. To improve efficiency and broaden test coverage, researchers have explored methods for automatically executing unit tests. 
The goal is to achieve comprehensive coverage of all possible scenarios while maintaining clarity and ease of understanding, thereby facilitating effective quality assurance and debugging processes \cite{yang2024empirical,fontes2023automated,jain2024testgenevalrealworldunit,yin2024getattentionbasedselfguidedautomatic,10438901}. 
This challenge extends to the evaluation of LLM-generated outputs, where assertions must validate not only functional correctness but also alignment with implicit requirements in natural language prompts, which is a core concern in prompt engineering and evaluation research \cite{10.1145/3654777.3676450}.

In the context of data wrangling, unit test generation can be more targeted. By extracting descriptive information about target scripts from user prompts, assertions can be set to verify script correctness \cite{yu2023improving}. 
However, in practice, users typically require direct validation of statistical properties (e.g., range ordering) rather than exact data comparisons. 
To enhance flexibility and structural integrity, the concept of constraint solving can be introduced into the test suite framework. 
Constraints are extensively utilized across various fields, including software engineering, visualization, and databases, to restrict the possible values of variables, thereby representing partial information about those variables \cite{Bartak1999}. 
In constraint programming, unknowns, such as variables, are expressed through a series of constraints, and any solution must satisfy all the restrictions of these constraints \cite{8440847}. 
Constraint programming standardizes knowledge representation, enabling users to model domain expertise as high-level expressions. 
% This approach facilitates a structured and efficient way to handle complex problems in different technical domains. 

LLMs have demonstrated potential in generating unit tests with high coverage \cite{yang2024empirical,chen2024chatunitest,wang2024hits,guilherme2023initial}. %Traditional unit test approaches often require precise expected output data for each step, which is possibly impractical \cite{10.1145/3583780.3614786,10298528}. Before deploying in specific scenarios, we might only have a rough idea of the expected outcomes, inferred from context.
When using an LLM agent, users usually provide a natural language prompt that both describes their requirements and acts as a specification for the desired script. 
This dual role of prompts necessitates methods to extract explicit and implicit conditions, echoing Shreya Shankar et al.'s work on aligning evaluations with human needs \cite{10.1145/3654777.3676450}, which can also be involved in the context of LLM-generated script debugging. 
%This prompt includes constraint information defining the solution space for executing specific data tasks \cite{10.1145/3638529.3654049, 10.1145/3587102.3588805, sandholm2024randomnessneedsemantictraversal}. The generated script must adhere to these constraints to ensure it meets basic functionality. 
An analysis program based on LLMs can extract the expected output criteria from the prompt \cite{zhu2023prompt, huang2024semanticguidedpromptorganizationuniversal}, which is then used to evaluate the functional correctness of the generated scripts. 
% A common practice in the field of machine learning is to use extracted constraint points for self-supervision of the generated results within the system \cite{gori2023machine,popescu2022overview,de2010constraint}. In contrast, we can expose these internal test points to users for transparency and customization. 
By comparing the actual outcomes with the test cases, the analysis ensures the scripts align with user specifications and performs as intended. 
This process enhances both the flexibility of the mechanisms and the alignment between scripts and user prompts.

\subsection{Advancing UIs for Human-LLM Interaction}
In the midst of rapid advances in LLMs, the HCI community has made significant strides in enhancing human-LLM interaction, surpassing common chatbot dialogues and basic API interactions.

Similar to our efforts to facilitate comprehension and verification of generated content, several studies focus on accelerating users' ability to understand and validate LLM outputs efficiently \cite{dong2024frameworkrealtimesafeguardingtext,10670515}. 
For instance, VIME \cite{10.1145/3654777.3676323} enables users to more efficiently understand and validate the outputs of sequential ML models. 
% Shreya Shankar et al. \cite{10.1145/3654777.3676450} specifically address the alignment of evaluation criteria with human needs. 
Our work centers on emerging scenarios where LLMs generate and debug data analysis scripts. 
We introduce features such as code scroll narration, streaming validation of code correctness post-generation, and step-by-step navigation through data transformation processes.

Additionally, other research explores innovative visual representations and interactive designs for LLMs beyond traditional single-text prompts. 
For example, Xie et al. \cite{Xie_2024} use interactive tree maps to assist in understanding. Advanced interactive applications using LLMs have been explored in areas such as educational analytics \cite{10.1145/3654777.3676347}, visualization generation \cite{yang2024matplotagentmethodevaluationllmbased} , and anxiety mitigation \cite{10.1145/3654777.3676437}. 
SHAPE-IT \cite{10.1145/3654777.3676348} highlights the ability to facilitate rapid ideation of a wide range of creative behaviors with AI. 
Following similar principles, our work enables users to interact with intermediate visualizations, allowing real-time refinement and modification of generated code. 
This provides a more intuitive and fine-grained control over LLM outputs, enhancing usability and effectiveness.
\section{Formative Study}
\label{section:3}
We conducted a formative study (N=8) to gain a better understanding of how users debug the LLM-generated data wrangling scripts and inform the design considerations. 
% better understand the methods for debugging the scripts generated by LLM in the data wrangling scenario and inform design considerations for contextualized support.
\subsection{Setup}
\textit{Recruitment \& Screening.} We posted recruitment advertisements on university forums and social media. Candidates were required to complete a questionnaire to provide their basic information and relevant work experience. We pre-screened volunteers who had experience using LLM-powered tools to generate data wrangling scripts for our interviews.

\textit{Participants.} We recruited 8 participants in total (P1-P8), with 2 females and 6 males, aged 20 to 30 years. Specifically, there were 4 postgraduate students whose daily research involves data analysis work (P1-P4), 2 undergraduate students majoring in Computer Science and Technology (P5-P6), and 2 data journalists who perform daily statistical reporting (P7-P8). 
All participants have prior experience in debugging data wrangling scripts generated by LLMs and correcting issues within these scripts. 
Given the varying levels of expertise they possess in programming skills and debugging such scripts, this allows them to provide rich and diverse perspectives from different angles.

\textit{Interview.} First, we asked participants to describe their experiences using LLMs to generate data wrangling scripts, including the specific scenarios in which they applied these tools. 
In addition, participants were requested to bring chat history from real work scenarios where they had requested LLMs to generate data wrangling code but found the output to be suboptimal. 
We explored how they identified issues within the scripts, performed diagnostic analyses, and ultimately corrected the errors. 
Each participant was compensated with \$10/hour. 
The formative interviews lasted 20-40 minutes. 
All interviews were audio-recorded. 

% \textit{Analysis.} Following thematic analysis methods, we first designed three key topics of inquiry. 
\textit{Analysis.} Following thematic analysis methods, we first set up three key topics of inquiry. 
These areas focused on the workflows, specific requirements, and common challenges faced by interviewees when debugging and modifying LLM-generated scripts.
To gather comprehensive insights, we combined participants' responses to a series of questions with information obtained through guided discussions during the interviews. 
By incorporating real-world examples and practical solutions from these discussions, we were able to systematically organize our findings and gain deeper insights. 
This approach enabled us to effectively guide the design of our system, ensuring it addresses the diverse needs and workflows identified through our analysis.

\subsection{Findings}
In the following, we present a thematic summary of the key points in the interview research results.

\subsubsection{For what tasks do users frequently rely on LLMs in generating data wrangling scripts, and how do these tools perform? }\ 

Participants recognized that, for simple tasks, conversational LLM agents offer a significant advantage in enhancing the efficiency of generating data wrangling scripts. They have experience using LLMs to generate such scripts across various task scenarios, including scientific research (6/8), routine learning tasks (4/8), machine learning training (2/8), business analytics (2/8), and investigative journalism (2/8). 
\textit{``Nowadays, whenever I face simple and highly repetitive data wrangling tasks, I opt to use LLMs to handle them swiftly. This allows me to allocate more time to other meaningful endeavors.''} (P7) 
With the assistance of LLMs, end users need only describe their task requirements to receive responses in the form of code along with functional reports. 
This feature has been particularly well received by users with less programming experience. 
\textit{``I have limited programming experience in data wrangling, and the LLMs allow me to obtain a usable script without learning programming specifically.''} (P6)

However, participants noted that, in complex scenarios, the effectiveness of conversational LLM agents in generating data wrangling scripts may not be promising. 
When the volume of requirements increases or tasks require deeper analysis, LLMs often produce scripts that either do not meet the task requirements or contain syntactic errors. 
% Participant P8, who has extensive experience in programming and algorithms, 
P8 stated, \textit{``When tasks require deep thinking, using LLMs often results in errors, which can make the efficiency of using them lower than if I were to write the code myself directly.''} 
Most participants (7/8) stressed that it is unrealistic to expect LLMs to generate correct and usable data wrangling scripts in a single round of dialogue in specific scenarios. 
% Instead, problem analysis and code refinement are necessary.
Instead, issue analysis and script refinement are necessary.

\subsubsection{How to identify issues in LLM-generated data wrangling scripts? }\ 
Generally speaking, after obtaining an LLM-generated script, the interviewees may go through the following two phases: First, they verify the script's usability. Second, upon discovering any issues, they have to assess the source of these problems. (\autoref{table:methods})

\begin{table}
    \caption{Methods for debugging LLM-generated data wrangling scripts and their limitations in practice.}
    \centering
    \tabcolsep=0.3cm
    \begin{tabular}{lp{4.2cm}}
    \toprule
        \textbf{Debugging Methods}&\multicolumn{1}{l}{\textbf{Main Limitations}}\\
    \midrule
        Initial Script Review&Experience-dependent and difficult to detect deep logic flaws.\\
        Test Dataset Execution&Inadequate edge case coverage and challenging validation.\\
        Segmented Print Output&Information overload and time-consuming for scope narrowing.\\
        Local IDE Debugging&High learning curve and complex environment setup.\\
        LLM-based Correction&May introduce new issues and is inefficient.\\
        % Script Modularization&Time-consuming to narrow down root cause scope.\\
    \bottomrule
    \label{table:methods}
    \end{tabular}
\end{table}

\textbf{Verify Usability.} After obtaining initial scripts, most participants (7/8) conduct preliminary syntax and format checks on the scripts output by LLMs, leveraging their experience to visually inspect for any obvious issues.
More experienced participants (4/8) can often determine whether the script meets the simple requirements by just reviewing it. 
However, they unanimously agreed that direct observation is not reliable and should only serve as a pre-processing step, requiring additional methods for thorough verification. 
% \textit{``I sometimes spot issues during the LLMs' streaming output and immediately terminate the generation to start a new conversation. This approach helps avoid initial errors, but it's far from sufficient.''} (P1) 
\textit{``While monitoring the LLM's streaming output, I sometimes terminate responses mid-generation when spotting issues and restart the conversation. Though this helps catch early issues, it remains an inadequate solution overall.''} (P1) 
Subsequently, participants may delve deeper through further dialogue with the LLM (3/8) or by switching to a local coding environment (5/8). 
The most straightforward method to test if a script is usable is to run it on a test dataset. 
Most participants (6/8) chose this approach to verify whether the script has compilation errors or fails to meet their requirements. 
However, P2 pointed out the data scale dilemma inherent in this method:  \textit{``When I test with a small dataset, it might not cover all edge cases of the specific scenario. When using a large dataset, the volume of output results becomes too heavy to easily validate correctness.''} 
% Moreover, participants (3/8) noted that they might not be able to fully determine the script's correctness just by examining the output results.
Participants (3/8) noted that they might not be able to fully determine the script's correctness just by examining the output results because of concerns about uncaught situations.

\textbf{Locate and Debug Issues.} Merely verifying the usability of a script may not suffice. 
Most participants (6/8) found that simply describing observed anomalies to LLMs often does not lead to accurate understanding and resolution; it can even misdirect efforts. 
\textit{``Categorizing the types of problems and narrowing down the scope of modifications greatly enhances the effectiveness of LLMs.''} (P5) 
% To reduce the effort of debugging, participants opted to analyze the root causes themselves and focus on specific code segments. During this process, outputting variable information during script execution proved invaluable for analysis. 
To reduce the effort of debugging, participants opted to analyze the source of issues themselves and focus on specific code segments. 
Throughout this process, reviewing variable information during script execution proved invaluable for analysis. 
Participants employed various methods, such as segmenting notebooks (5/8), directly printing outputs (6/8), and using IDEs to inspect register values (3/8), to trace the process. 
% \textit{``In Python, printing the first five rows of a table is a common and efficient way to preliminarily verify if the script functions correctly, but this method might miss issues in subsequent rows.''} (P2) 
\textit{``In Python, printing several rows of a table is a common and efficient way to preliminarily verify if the script functions correctly, but this method might miss issues in subsequent rows.''} (P2) 
Furthermore, scripts generated by LLMs may not align with users' coding preferences. 
Participants (2/8) mentioned encountering overly complex single-step code generated by LLMs, which complicates debugging. 
Most participants (7/8) agreed that simpler, clearly segmented code is much easier to debug.

\subsubsection{What challenges have been encountered when resolving issues in LLM-generated data wrangling scripts? }\ 

Three themes emerged when participants described their own difficulties in modifying the scripts.

\textbf{Debugging by fully checking output data is difficult and tedious. }Participants (6/8) pointed out that executing scripts on a test dataset and then checking the output data for debugging is a practical approach. 
But the manual assess methods mentioned above place high demands on users' capabilities. 
Some participants (4/8) noted that verifying the correctness of data transformations and ensuring that the generated scripts produce accurate results are tedious and error-prone tasks. 
\textit{``Checking the output (data results) from the generated script cell by cell is cumbersome.''} (P4) 
To address this, participants (5/8) suggested aggregating data results (e.g., calculating the mean and median of a column of data) or print out a small sample of the data (e.g., the first five rows of a data table) during the execution process to form reports as a way to reduce complexity and alleviate the burden caused by large data scales. 
% However, some participants (3/8) expressed concerns that this approach might disrupt the inherent structure of the data, so it does not reliable when assisting in script modification.
However, some participants (3/8) expressed concerns that this approach does not have enough representation to describe data entirely, so it does not reliable when assisting in script modification.

\textbf{Modifying hampered by limited programming skills. }Although LLMs can provide clear annotations to explain each step, participants (8/8) still found modifying the generated code challenging. On one hand, LLMs may use unfamiliar packages or advanced syntax, making adjustments difficult. 
\textit{``Sometimes there are functions I don't recognize or long statements with deeply nested structures, and even after consulting documentation, I still can't figure out how to correct them.''} (P3) 
On the other hand, addressing issues introduced by LLMs often requires both meticulous attention and programming experience. 
\textit{``It's common to misjudge the source of a problem, which leads to incorrect modifications, wasting time and effort and potentially causing additional issues.''} (P1) 
% Even with LLMs, developers still struggle with making effective changes, particularly when dealing with intricate logic or unfamiliar constructs.
Even with LLMs, developers still struggle with making effective modifications, particularly when dealing with intricate logic or unfamiliar constructs.

\textbf{Refining iteratively and identifying hard-to-detect hallucinations. }Most participants (7/8) also opt to engage in further dialogue with LLMs to iteratively refine the script, when manual and simple modifications are insufficient. 
Unfortunately, this process requires significant effort to communicate the nuances of the desired analysis outcomes and carefully avoid introducing new errors. 
% In the scenario of introducing multiple rounds of dialogue, participants (3/8) have encountered instances where LLMs amplify hallucinations, leading to an increase in script errors. 
In the scenario of introducing iterative dialogue, participants (3/8) have encountered instances where LLMs amplify hallucinations, leading to an increase in script errors. 
\textit{``The agents may misinterpret my intended data wrangling steps that are not syntactically valid, or suggest using functions that do not exist in the given context. 
These errors sometimes accumulate recursively and are difficult to detect.''} (P6) 
Consequently, most participants (6/8) are reluctant to fully rely on LLM-generated results and also use local environments for debugging. 
However, even with this approach, they still often find themselves caught in iterative cycles, spending more time identifying and resolving emerging issues, or having to abandon the current conversation history and start a new dialogue.

\subsection{Design Considerations}
In our formative study, all participants expressed significant interest in incorporating LLMs into their workflow to generate and debug data wrangling scripts. 
% In our formative study, participants are currently using LLMs code agents to generate data wrangling scripts, but they have all expressed significant interest in enhancing the experience of debugging these scripts within their workflows. 
Our design goal is to help users quickly verify the reliability, identify issues, and correct discrepancies in the LLM-generated data wrangling scripts. 
To achieve this, we propose the following design considerations (D1-D5) to guide the design of the system.

\textbf{D1. Automatically generate scripts and corresponding test cases for reliability detection based on natural language prompts. }After issuing requests to LLMs, there is a risk that the generated data wrangling scripts may contain issues, so debugging these scripts is a necessary step. 
% \textit{``Having a ready tool to assist in debugging after generating the script (with LLMs) would greatly help maintain task coherence and improve work quality.''} (P4)
\textit{``Having an integrated tool to assist in debugging after generating the script (with LLMs) would greatly help maintain task coherence and improve work quality.''} (P4)
Automatically translating implicit information within natural language prompts into test cases for reliability detection can significantly enhance work efficiency.

\textbf{D2. Execute unit tests, and provide an overview of test results and data transformation progress with intuitive visualization. }Unit testing is an effective and low-threshold method to ensure the reliability of scripts. 
The model should be designed to perform systematic and high coverage verification. 
However, even after extracting test points from the prompts and performing automated detection, understanding test information in constraint form and interpreting complex test results can still be challenging for users.
To address this, a visualization approach can be introduced, aimed at intuitively summarizing test outcomes. 
This method should be designed to effectively present structured, step-by-step information, aiding users in comprehending data transformations within scripts. 
By making error propagation more transparent, it helps users identify the key points in the execution process where failures occur, ultimately supporting more efficient debugging.

\textbf{D3. Allow users to examine the details of the data for a specific step and customize the test set. }In addition to providing an overview of test results, the system should also support viewing and modifying detailed information at any stage. 
Examining the data details of a specific step in the script, similar to breakpoint debugging, is a common practice in existing workflows and is highly beneficial for analyzing the issues. 
Furthermore, based on our automatically generated test cases, users may want to make additional customized modifications to support further testing. 
The system needs to allow for adding or modifying existing test cases.

\textbf{D4. Support script modification based on test reports. }Based on our formative study, we found that users often choose to send both the compiler error messages and the script to LLMs during local compilation and debugging. 
This method is straightforward and effective. 
Including test results in the prompt provides the LLM with detailed insight into the problem. 
These contexts, such as error messages, help the LLM better understand the issues and identify the root causes, allowing for more targeted and precise suggestions to solve the problems.
Inspired by this workflow, in our testing framework, if users want to debug a script, we should allow users to send the test results as part of the prompt to the LLMs for analysis and problem resolution, in addition to performing manual modifications and providing more natural language descriptions about their requirements.

\textbf{D5. Integrate with existing data wrangling workflows seamlessly. }Current approaches often lead to a disconnected workflow that disrupts the natural flow of data wrangling tasks. 
For instance, users may need to switch between different platforms to debug code, which can cause mental context-switching and break the continuity of their work. 
Our design tries to reduce these issues by keeping everything within a cohesive workflow.
We can embed elements directly within the response box in the conversational user interface (CUI), allowing users to stay focused on their script and analysis without unnecessary interruptions.
To ensure seamless integration with the existing design, our visualization elements and interaction designs are embedded within the response box, maintaining logical consistency with the original output information. 
% Additionally, the approach can also keep low-code or even no-code environments of LLMs for analysis and modification, further improving accessibility and user experience.
% \input{sections/4_usage_scenario}
\section{ViseGPT}
\label{section:5}

\begin{figure*}[htb]
  \centering
  \includegraphics[width=2\columnwidth]{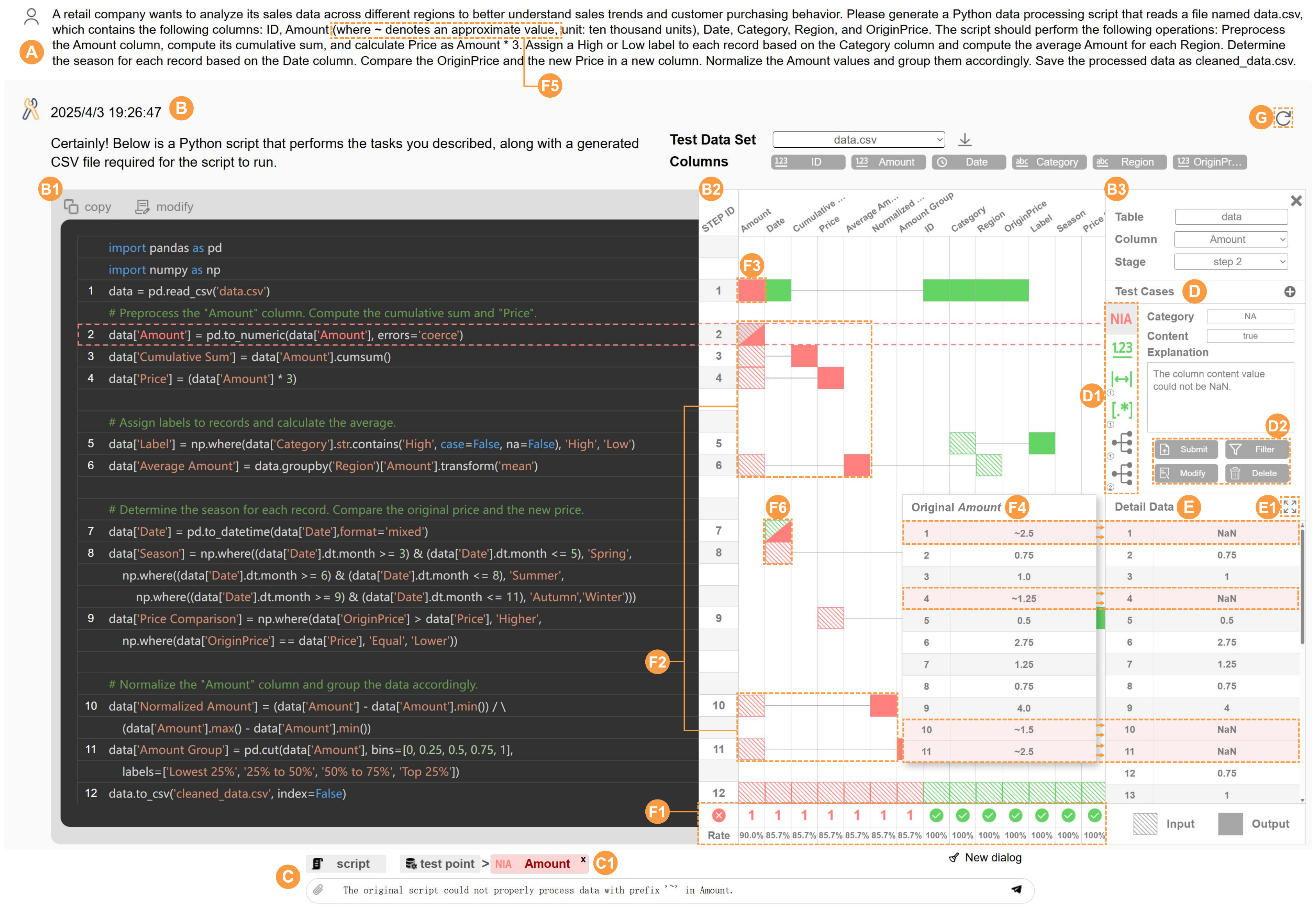}
  \caption{A screenshot of the ViseGPT user interface. 
  After users send a prompt to ViseGPT (A), the response content in the reply box (B) is organized into three components: 
  (B1) A script view similar to those in common LLM-based code agents;
  (B2) A tailored Gantt chart providing an overview of test results;
  (B3) A detail view displaying the test case list and detailed data for one column at one step.
  Users can resend prompts for script iteration (C), with the option to attach test reports (C1).
  }
  \Description{A detailed workflow diagram showing the ViseGPT process, including steps A, B and C.}
  \label{fig:system}
\end{figure*}

% Based on our interviews, users need a efficient and low-barrier approach to improving the debugging of LLM-generated data wrangling scripts. 
% ViseGPT can automatically extract constraints on script output data from natural language prompts , process them into test cases (\textbf{D1}) and conducting unit testing (\textbf{D2}).
% To integrate seamlessly with existing data wrangling workflows (\textbf{D5}), ViseGPT supports the customization of unit tests and the presentation of data details (\textbf{D3}). 
% Once the test results are obtained, test reports can be sent as an enhanced prompt to the LLM code agent, completing the entire workflow loop (\textbf{D4}). 
% All in all, the design of ViseGPT consists of three main parts: generating and executing test cases, visualizing unit test flow, and optimizing the script iteration workflow. 
Based on user interviews, ViseGPT should address users' need for an efficient, low-barrier approach to debugging LLM-generated data wrangling scripts. The system comprises three core components: (1) \textit{Automated Test Generation and Execution}: ViseGPT automatically extracts output constraints from natural language prompts, processes them into test cases (\textbf{D1}), and conducts unit testing (\textbf{D2}); 
(2) \textit{Workflow Integration and Visualization}: To ensure compatibility with existing data wrangling workflows (\textbf{D5}), the system supports unit test customization and provides detailed data visualization (\textbf{D3});
(3) \textit{Closed-Loop Iteration}: Test reports are automatically repurposed as enhanced prompts for the LLM code agent, forming an iterative debugging loop (\textbf{D4}).

% \section{Usage Scenario}
\subsection{Usage Scenario}

Mary, a marketing specialist at a retail company, intended to 
% To gain deeper insights into product sales trends, she intended to 
analyze sales data across various quarters and regions after processing the original dataset. 
% For this purpose, she turned to \textit{ViseGPT}, a conversational data wrangling script debugging tool powered by LLMs that she is well-acquainted with.
% While directly submitting a prompt to LLM could quickly generate a script, she was unsure whether the output would fully meet her requirements—so she turned to \textit{ViseGPT}, a conversational debugging tool for LLM-generated data wrangling scripts. 
% Mary uploaded a natural language description of her data wrangling task, specifying the key requirements to preprocess the existing data and extract basic market data metrics.
Although LLMs offer rapid script generation capabilities, the reliability of their output for precise data wrangling requirements remained a concern.
She opted to use \textit{ViseGPT}, a conversational debugging tool designed for LLM-generated data wrangling scripts.
After describing her data wrangling requirements in natural language, \textit{ViseGPT} generated a Python script and executed comprehensive unit tests, displaying the results in the tailored Gantt chart (\autoref{fig:system} B2). 
In the Gantt chart, each column represented an output data column, while each row aligned with the output script step. 
% Mary initiated a conversational session to specify the data wrangling requirements in natural language.
% In response, ViseGPT generated a script to execute the data wrangling task. 
% In addition to providing the code and basic informational responses, the system performed unit tests on the generated script and embedded dynamic visualizations of the test results on the right side of the response panel.
% In response, \textit{ViseGPT} generated a Python script and automatically performed unit tests, with the test results displayed on the right panel (\autoref{fig:system}).
By observing the summary rows at the bottom (\autoref{fig:system} F1), Mary found that some columns in the output data failed to meet the testing criteria, so she decided to inspect the script in detail to resolve the issues.

\textbf{Exception issues due to type coercion.} Mary first reviewed the tailored Gantt chart (\autoref{fig:system} B2)  for the script. 
She identified a series of issues that could be traced back to the \texttt{Amount} column in the overview (\autoref{fig:system} F2), which subsequently propagated through to the \texttt{Cumulative Sum}, \texttt{Price}, \texttt{Price Comparison}, and \texttt{Normalized Amount} columns. 
To address the root cause, she initially focused on the \texttt{Amount} column in Step 1 by clicking on its corresponding rectangle in the Gantt chart (\autoref{fig:system} F3). 
Upon inspection, she found that the input data had type issues. 
After applying a filter, Mary discovered that all the data failing the tests contained the “\textasciitilde” prefix in the \texttt{Amount} (\autoref{fig:system} F4), used to denote approximate values (\autoref{fig:system} F5). 
This format was incompatible with the float type expected in subsequent operations.
However, directly forcing type conversion could result in invalid values.
Therefore, it was necessary to strip prefixes while retaining the numerical value (for example, converting “\textasciitilde50” to “50”).

Then, in Step 2, she observed that the output did not satisfy the test case of having no missing values in the column (\autoref{fig:system} E). 
It became evident that the initial failure to properly handle the non-float formatted data led to anomalies during type coercion in \texttt{pd.to\_numeric(data['Amount'])}, resulting in missing values. 
% These anomalies, in turn, impacted subsequent steps involving the \texttt{Amount} column, such as calculations for \texttt{Cumulative Sum} and \texttt{Price}, thereby producing incorrect results.
The missing values subsequently propagated through dependent calculations, affecting the \texttt{Cumulative Sum} and \texttt{Price} computations with incorrect results.

After identifying the source of the issues, Mary sent ViseGPT the relevant information (\autoref{fig:system} C), including details about the failing test case (\autoref{fig:system} C1) and the requirement to remove the prefix while retaining the floating-point numbers.
In the newly generated script, preprocessing steps were added to handle the format of the \texttt{Amount}, resulting in a reduction in the number of test issues.

\begin{figure}[H]
  \centering
  \includegraphics[width=\linewidth]{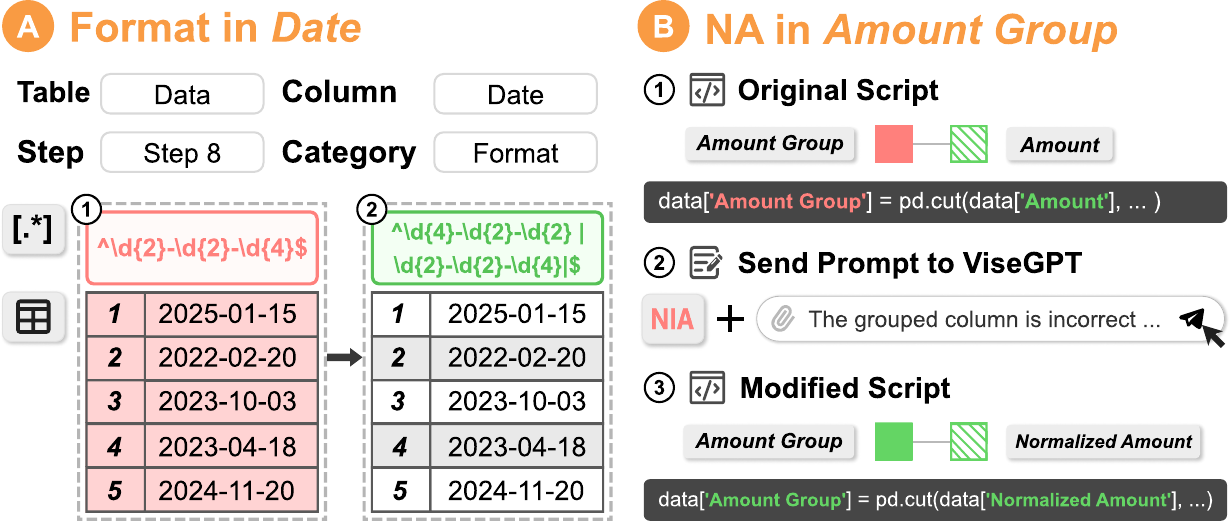}
  \caption{Two issues in the usage scenario. (A) Format issues that require expanding the test case. (B) Exception issues arising from grouping by incorrect column.}
  \Description{Two issues.}
  \label{fig:scenario}
\end{figure}

\textbf{Format issues that require expanding the test case.} Subsequently, Mary started analyzing the \texttt{Date} column issues identified in Steps 8 and 9 (\autoref{fig:system} F6).
By clicking on the rectangles corresponding to these steps, she examined the list of test cases in the detail view and identified that the errors were due to format mismatches during validation (\autoref{fig:scenario} A).

Upon reviewing the prompt, Mary realized that she had not explicitly specified the expected date format for the \texttt{Date}. 
As a result, the automatically generated test cases used  `\textbackslash d\{2\}-\textbackslash d\{2\}-\textbackslash d\{4\}' as the default format for validation (\autoref{fig:scenario} A1). 
However, when the data was processed in \texttt{pd.to\_datetime(data['Date'],format='mi- \newline xed')}, it was converted to the `\textbackslash d\{4\}-\textbackslash d\{2\}-\textbackslash d\{2\}' format, causing discrepancies between the test cases and the processed data. 
To resolve this issue, Mary directly modified the test case content to align with the `\textbackslash d\{4\}-\textbackslash d\{2\}-\textbackslash d\{2\}' format (\autoref{fig:scenario} A2). 
Then she reran the tests (\autoref{fig:system} G) and confirmed that the issues were resolved.

\textbf{Group label issues arising from grouping by incorrect column.} After adjusting the test cases and rerunning them, Mary turned her attention to the issues in the \texttt{Amount Group} column in Step 12 (\autoref{fig:scenario} B). 
This column was supposed to be a label column that categorized the normalized \texttt{Amount} values into groups based on their magnitude. 
However, when examining the tailored Gantt chart, she noticed that the grouping was incorrectly being performed on the original \texttt{Amount} column instead of the \texttt{Normalized Amount} (\autoref{fig:scenario} B1).

Upon inspecting the specifics of the \texttt{Amount Group} in this step, Mary found that it contained missing values. 
To investigate further, she clicked the magnify
button (\autoref{fig:system} E1) in the top-right corner of the detail data view for a more detailed look. 
She found that the original \texttt{Amount} column was being used as the basis for grouping, which caused most values — that is, those greater than 1 — to fall outside the expected normalized range. 
As a result, these values could not be properly grouped, leading to missing values in the \texttt{Amount Group} column.
To address this issue, Mary uploaded the test results and the requirement to ViseGPT (\autoref{fig:scenario} B2). 
Following her submission, the grouping column was correctly modified to \texttt{Normalized Amount} (\autoref{fig:scenario} B3). Finally, the script passed all the test cases, and Mary got a script that exactly matched her prompt.

\begin{figure*}
  \centering
  \includegraphics[width=1.95\columnwidth]{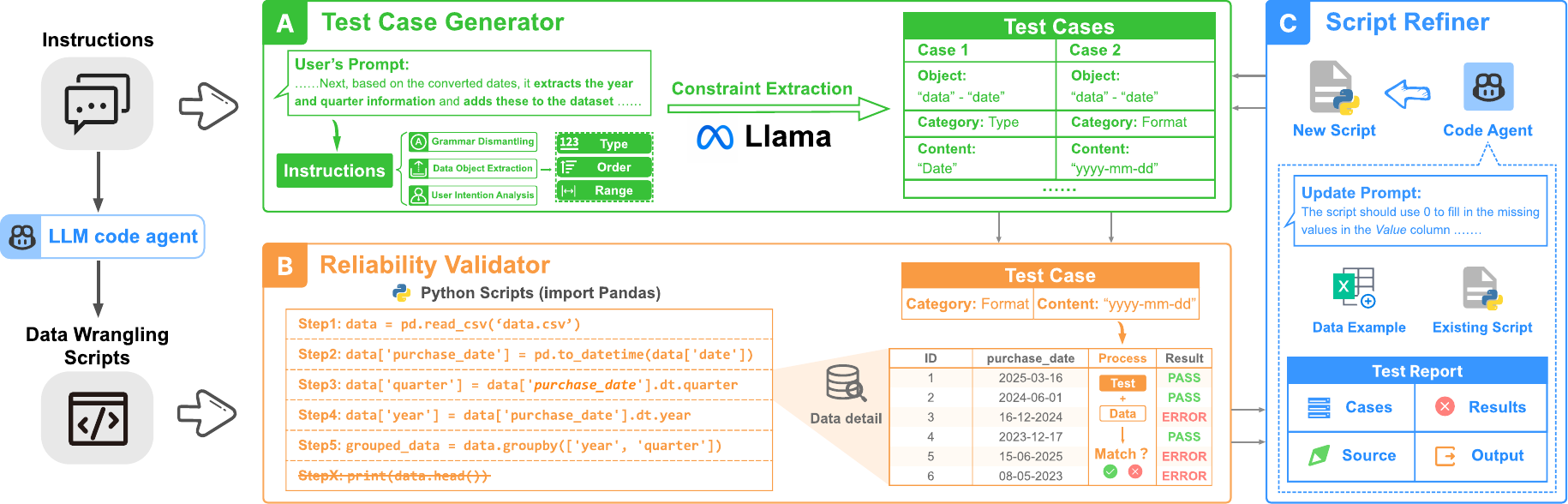}
  \caption{ViseGPT Workflow. (A) Test Case Generatorgenerates test cases based on data constraints extracted from users' prompts; (B) Reliability Validator executes the data wrangling scripts generated by the LLMs code agent, matches the output data results with test cases, and produces test reports; (C) Script Refiner can iteratively modify scripts by sending test reports, natural language instructions, and data examples as prompts to LLMs.}
  \Description{A detailed workflow diagram showing the ViseGPT process, including steps A, B and C.}
\end{figure*}

\subsection{Generating and Executing Test Cases}
\label{section:5.1}
% Before the formal introduction of the design, it is necessary to follow up on the conclusions from the previous interview
% Based on our interviews, users need a highly efficient and low-barrier approach to improving the debugging of LLMs-generated data wrangling scripts. 
With the assistance of LLM, ViseGPT can automatically derive output constraints from natural language prompts and processes them into test cases for unit test execution. 
In another work on data wrangling scripts, Ferry \cite{10670464}, there is already a relatively comprehensive summary of the classification of data constraints. 
Building on this, besides \textbf{Type}, \textbf{Format}, \textbf{Range}, \textbf{Order} and \textbf{Exception} (i.e., \textbf{Missing Value} in Ferry), ViseGPT's classification framework has also undergone certain adjustments:

% Although numerous studies on data constraints exist, they do not fully address the specific needs of tables in the data wrangling domain or lack sufficient generality for practical scenarios.  
% To clarify what types of constraints can be extracted from natural language prompts, whether these constraints can be utilized as test points, and if providing this information meets requirement, we conducted an extensive review of existing work and analyzed materials obtained from our previous formative study interviews. 
% This process led us to identify the following eight constraint types that can be used for unit testing the output data results from scripts:
\begin{itemize}[leftmargin=10pt]
% \item \textbf{Data types} of columns are categorized into six basic categories: Logical, Integer, Float, Character, Date, and Time. Some of the following constraints vary based on the data type.
\item \textbf{Unique} verifies that each value in the column appears only once (e.g., an ID column used as a primary key).
% \item \textbf{Range} verifies the value ranges for Integer/Float data and the string length ranges for Character data. 
% Notably, we can use the $\cup$ (union) and $\cap$ (intersection) symbols to express multi-segment, non-contiguous range constraints (e.g., $(a, b] \cup [c, d]$, or $(a, b) \cap [c, d)$).
% \item \textbf{Format} specifies the text format of data values. We use regular expressions and format patterns for static detection (e.g., 'yyyy-mm-dd'). Moreover, we can generate natural language descriptions of these formats to enable more flexible testing.
% \item \textbf{Order} specifies the sorting of data values (e.g., ascending, or descending).
\item \textbf{Forbidden Value} requires that certain specific values must not appear in the data.
% \item \textbf{Exception} verifies the presence of outliers (e.g., NA).
\item \textbf{Relation} describes the dependencies between columns and is further elaborated based on the different data types involved. For example, common equality relationships, magnitude relationships for Integer/Float data (e.g., greater than/less than), and substring/superstring relationships for Character data. Besides these basic relation categories, we enable the use of natural language labels to flexibly describe more complex dependencies.
\end{itemize}
Under the aforementioned constraint classification framework, the analysis program can extract user requirements from the prompt for subsequently testing the generated script.

% To evaluate the correctness of data wrangling scripts generated by LLMs within our constraint framework and to demonstrate the changes in data correctness throughout the script flow, it is necessary to dissect the scripts step-by-step. 
To evaluate the correctness of data wrangling scripts generated by LLMs within our constraint framework and to demonstrate the changes in data correctness throughout the script flow, it is necessary to separate the scripts step-by-step. 
Then ViseGPT generates test input datasets that meet the prompt requirements and script variable parameters, and systematically match output data results with test cases. 
Finally, ViseGPT summarizes the test results.

\textit{Separating the Scripts Step-by-Step.} 
% Typically, the logical environment constructed by data processing scripts is chained stepwise (e.g., pandas in Python, or R). 
% According to the syntactic logic, 
% ViseGPT decomposes a script into a series of atomic steps forming a chain and remove any redundant or unnecessary repetitive steps. 
After sending the user's natural language instructions to LLM embedded in ViseGPT, the system generates the corresponding data wrangling script. 
ViseGPT can then break down the script into a series of atomic steps forming a chain and ignore any redundant or unnecessary repetitive steps (e.g., steps like print and assert that do not affect the data).
% We extract the input-output elements (i.e., tables and columns) from the atomic steps.
The system will use AST parsing based on Python syntax to extract the input-output elements (i.e., tables and columns) from each atomic step, aggregate and collect them for subsequent visualization.

% Furthermore, to ensure the usability of the generated data, we name the test datasets according to the data object names obtained during the step-by-step script parsing process. 
% This approach ensures a clear identification and purpose of each dataset, facilitating subsequent data processing and validation efforts.
% \vspace{1em}

\textit{Constraint Matching Testing.} 
% We execute test datasets through LLM-generated data wrangling scripts to verify if each step's output meets the specified constraint test points. 
ViseGPT executes test datasets through LLM-generated data wrangling scripts to verify if each step's output meets the specified constraint test cases. 
Users can choose to upload their own test datasets, and the system will also generate a dataset that aligns as closely as possible with the task description based on the user's prompt. 
During the testing process, a static match will first be conducted to determine whether the output data follows the basic test cases.
For instance, during Type inspection, regular expressions can be utilized to ensure that the data format matches specified characteristics; for Range testing, checks are performed to ascertain whether numerical values or string lengths fall within designated intervals. 
To achieve sufficient coverage, ViseGPT also supports testing with LLMs in the categories of \textbf{Format}, \textbf{Forbidden Value} and \textbf{Relation}. 
Besides employing static matches, we also leverage natural language descriptions for constraints, utilizing LLMs to assess whether the output data conforms to these described requirements.

\subsection{Visualizing Unit Test Flow}
\label{section:5.2}
After test execution, in order to create an intuitive overview of the test report for locating the source of issues and assisting debugging, it is essential to use an appropriate method to visualize the script's data transformation process. 
Our design considered table and tree alternatives that leverage different structures to encode data flows, but the feedback received in iterative progress showed that tables obscure temporal dependencies while trees lack intuitive flow representation. 
In contrast, Gantt charts offer a stronger balance of information density, temporal clarity, and workflow mapping, with particular advantages in summarizing process information \cite{ramachandran2019gantt}, making them well-suited to our application scenario.
% It's known that the Gantt chart has advantages in summarizing process progress information \cite{ramachandran2019gantt}, making it well-suited to our application scenario.
Hence, a design based on Gantt charts is introduced in ViseGPT, following a user-centered development process that has been iteratively refined. 
In this framework, shape and color coding are used to generalize the test results, with an ``aligned-to-line'' design that segments scripts at the granularity of individual steps to align Gantt chart rows with LLM outputs side by side, helping users validate generated code.

\textit{Visualization of the data transformation process.} 
In Section \ref{section:5.1}, during unit test execution, ViseGPT has already separated the scripts step-by-step, simplifying the script content and extracting the main operations that impact the test results. 
% Next, we use a Gantt chart to summarize the script’s content.
Next, a tailored Gantt chart is applied in ViseGPT to visualize the script execution process, facilitating clear presentation of test results.
The step ID is shown on the left, aligning with the corresponding lines of scripts to create an intuitive visual mapping.
Specifically, we represent the data columns or global variables involved in each step as rectangles, using different legends to indicate whether a variable is an input or output. 
% Variables appearing in the same step are connected by lines, ensuring alignment with the corresponding lines of script. 
Variables appearing in the same step are connected by lines, allowing users to quickly locate when issues arise. 
% This design is primarily intended to effectively summarize script information (\textbf{D2}) and naturally complement the existing, yet incomplete, debugging workflow (\textbf{D5}).
The chart is arranged from top to bottom, aligning with the execution flow of the script. 
ViseGPT focuses on sequential execution and does not support branching or loops. 
This design aligns with typical data wrangling scripts, where batch operations are inherently linear and account for the majority of transformations. 
In the future, support for nonlinear logic can be expanded. 
In order to save space, column and variable names at the top are displayed at an angle. 
While step-based granularity ensures alignment with scripts, we acknowledge there are scalability challenges for long scripts and believe exploring hierarchical block-level abstractions and more detailed single-step interaction is meaningful (Section \ref{section:7.3}).

\textit{Visualization of unit test results.} Building on the visualization of the script’s data transformation process, the system can also present the execution process and final results of unit testing. 
Different rectangle colors indicate the test pass status of a variable at each step: green represents that the script’s output fully meets the constraints of the test cases, while red indicates conflicts. 
At the bottom of the summary chart, the final test results are displayed (\autoref{fig:system} F1), including the number of failed test cases and the test pass rate (the number of passed test cases divided by the total number of test cases). 
Users can click on the rectangles to view details for a variable at a specific step and customize the test set (\textbf{D3}). 
The detail view that pops up on the right includes variable parameters, step ID, test cases, and detailed data. 
When a rectangle is clicked, the corresponding script line and chart row will be emphasized, and the variable name in the script will be highlighted.

\subsection{Optimizing the Script Iteration Workflow}
After presenting the overview of test results, ViseGPT provides the corresponding features when users want to view the test cases and data details to make revisions for continued iteration.

\textit{Presentation and customization of test cases. }A detailed list of test cases for the corresponding variable is placed in the detail panel (\autoref{fig:system} D). 
The icons on the left side of the list represent different categories, while their colors indicate whether the test case has passed (\autoref{fig:system} D1). 
Clicking an icon reveals detailed information on the right, including a brief description and an explanation about the test case. 
Below the test case content, there are four buttons that enable custom modifications, sending test reports, and filtering conflicting data, enhancing the debugging and script analysis capabilities (\autoref{fig:system} D2). 
After customizing the test cases, clicking the refresh button (\autoref{fig:system} G) at the top will trigger the system to re-run the tests and display an updated test report.

\textit{Inspection of single-step detailed data.} Below the detailed list of test cases, the default view displays the output of the column at a specific step (\autoref{fig:system} E). 
After applying the data filter in the test case list, only the data associated with failed test cases will be shown.
However, inspecting a single column in isolation is not intuitive in complex scenarios involving multi-column and multi-variable transformations. 
For seeing several columns at the same time, users can click the magnify button in the upper right corner (\autoref{fig:system} E1) to view the entire data tables for the current step.

\textit{Iteration of the script based on test reports.} In daily programming tasks, when encountering system errors while working in a local compiler, it has become a common practice to directly copy the error message and send it to LLMs for analysis and resolution. 
ViseGPT supports sending test reports as prompts for code iteration (\autoref{fig:system} C1). 
The test report uploaded to the backend includes test case details, test results, script lines, and output data, allowing for more targeted problem-solving within our testing framework.

\subsection{Implementation}
ViseGPT is a web application, whose front-end is built on the React \cite{meta2024react} framework with TypeScript. 
The back-end utilizes the Llama-3.3-70b-Instruct model hosted by NVIDIA.
Based on this model, we have further developed the test case generator. 
Note that we select this model as a feasible solution, and the system’s tooling is largely model-agnostic, designed to work with various LLM outputs. 
We encourage future research to investigate alternative LLM architectures and their impacts on performance in depth. 
After the LLM code agent generates a data wrangling script, the system backend parses the prompt to analyze its semantics and processes it into test cases. 
% For efficient data management and handling interactions, we have also deployed a Python backend server implemented with Flask \cite{pallets2024flask}. 
The backend with WebSockets \cite{lpinca2025websockets} can progressively stream output to the frontend, allowing script content and test results to be dynamically displayed in real-time.
\begin{table*}
    \caption{The success rate (\%) and No. query submissions to LLMs in ViseGPT and Baseline for Task A-D (N=9/condition). 
    The issue column describes the mistake made by LLMs in the task. 
    \#CR: No. clicking rectangles in ViseGPT. 
    \#QT: No. query submissions with test reports in ViseGPT. 
    “(Value)”: standard deviance.}
    \renewcommand{\arraystretch}{1.2} 
    \tabcolsep=0.3cm
    \begin{tabular}{clcc|p{2cm}|p{2cm}|p{2cm}|p{2cm}}
        \Xhline{0.85pt}
        \multirow{2}{*}{\textbf{Task}} & \multirow{2}{*}{\textbf{Issue}} & \multirow{2}{*}{\textbf{\#CR}}& \multirow{2}{*}{\textbf{\#QT}}& \multicolumn{2}{c|}{\textbf{Average Query}} & \multicolumn{2}{c}{\textbf{Success (\%)}} \\
        \cline{5-8} &&&&\multicolumn{1}{c|}{ViseGPT} & \multicolumn{1}{c|}{Baseline} & \multicolumn{1}{c|}{ViseGPT} & \multicolumn{1}{c}{Baseline} \\
        
        \Xhline{0.6pt}
        \textbf{A}&Length mismatch in string&7.56 (6.45)&1.33 (0.67)&\multicolumn{1}{r|}{2.00 (0.94)}&\multicolumn{1}{r|}{3.33 (2.26)}&\multicolumn{1}{r|}{100 (0.00)}&\multicolumn{1}{r}{78 (0.42)}\\
        \textbf{B}&Missing boundary in grouping&3.89 (3.38)&1.33 (0.47)&\multicolumn{1}{r|}{1.78 (1.03)}&\multicolumn{1}{r|}{4.11 (1.97)}&\multicolumn{1}{r|}{100 (0.00)}&\multicolumn{1}{r}{33 (0.47)}\\
        \textbf{C}&Inconsistent letter casing&9.67 (8.38)&1.78 (0.79)&\multicolumn{1}{r|}{2.33 (0.67)}&\multicolumn{1}{r|}{3.11 (1.97)}&\multicolumn{1}{r|}{56 (0.50)}&\multicolumn{1}{r}{22 (0.42)}\\
        \textbf{D}&Non-compliant rounding&8.89 (5.07)&2.00 (0.94)&\multicolumn{1}{r|}{3.00 (1.63)}&\multicolumn{1}{r|}{2.78 (1.75)}&\multicolumn{1}{r|}{89 (0.31)}&\multicolumn{1}{r}{11 (0.31)}\\
        \Xhline{0.85pt}
    \end{tabular}
    \label{table:result}
\end{table*}

% \begin{figure}[h]
%   \centering
%   \includegraphics[width=\linewidth]{figures/time.pdf}
%   \caption{Spent time completing four tasks.}
%   \Description{Spend time.}
%   \label{fig:time}
% \end{figure}

% \begin{figure}[h]
%   \centering
%   \includegraphics[width=\linewidth]{figures/UEQ.pdf}
%   \caption{Rating results on Baseline and ViseGPT in UEQ.}
%   \Description{UEQ.}
%   \label{fig:UEQ}
% \end{figure}

\begin{figure*}[t]
  \centering
  \begin{minipage}[t]{0.48\textwidth}
    \includegraphics[width=\linewidth]{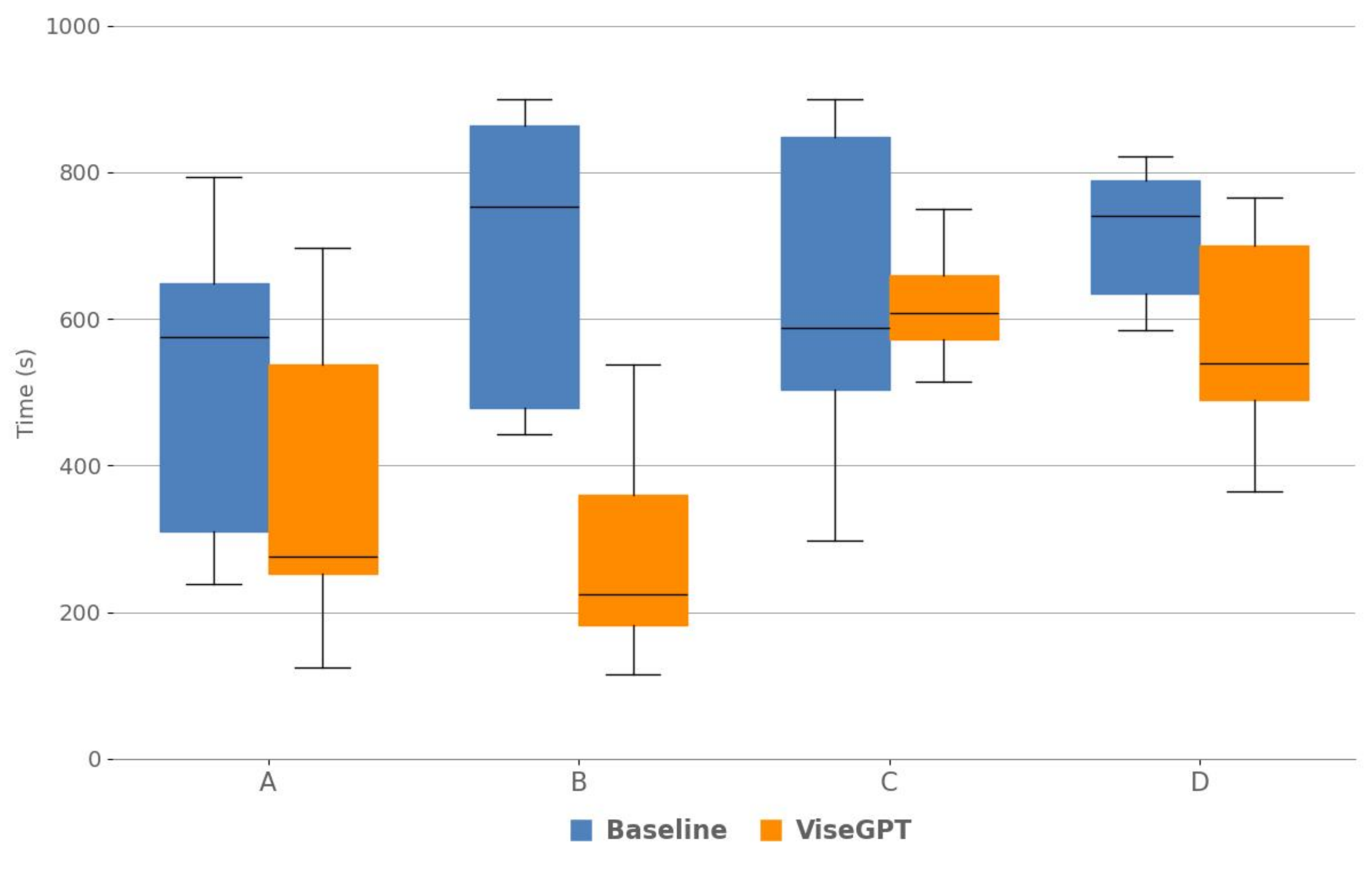}
    \caption{Completion time in four tasks.}
    \label{fig:time}
  \end{minipage}
  \hfill
  \begin{minipage}[t]{0.48\textwidth}
    \includegraphics[width=\linewidth]{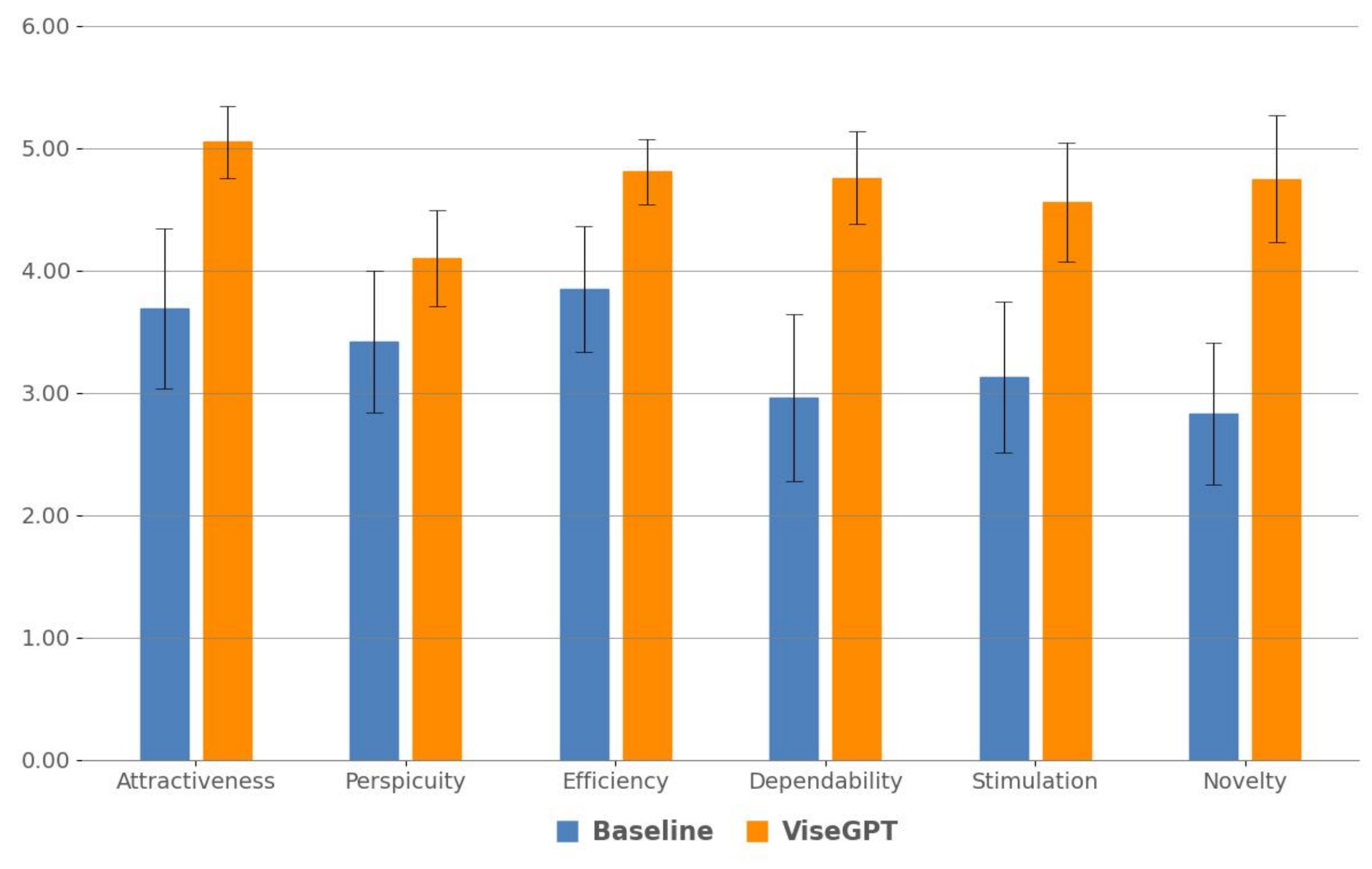}
    \caption{Rating results in UEQ.}
    \label{fig:UEQ}
  \end{minipage}
\end{figure*}

\section{Evaluation}

We evaluate ViseGPT through a user study with 18 participants (P1-18) of various backgrounds. The evaluation explores the following four research questions.

\begin{itemize}[leftmargin=10pt]
\item How does ViseGPT improve the efficiency of debugging data wrangling scripts generated by LLMs?
\item What impact does ViseGPT have on users' accuracy in evaluating LLM-generated data scripts?
% To what extent does ViseGPT enhance users’ accuracy in judging LLM-generated data wrangling script correctness?
\item How does ViseGPT assist users in iterating on LLM-generated data wrangling scripts after identifying issues?
\item How can ViseGPT be optimized to enhance its practical utility for debugging LLM-generated data wrangling scripts?
\end{itemize}

\subsection{Experiment Settings}
We conducted a comparative study with 18 participants to debug four LLM-generated data wrangling scripts between ViseGPT and Baseline. 
After completing tasks, we analyzed their performance and collected subjective feedback with the within-subjects design.
% In the experiment, 18 participants were invited to debug four LLM-generated data wrangling scripts using both ViseGPT and Baseline. Their performance and subjective evaluations were then analyzed.

\textbf{Participants. }We recruited 18 participants (15 males, 3 females; ages 20-26, M = 22.17, SD = 1.69) through campus forums, social media and promotional referrals. 
Participants comprised 12 undergraduates, 4 master students, and 2 PhD students  majoring in computer science, robotics engineering, bioinformatics, etc.
% Their proficiency in writing data wrangling scripts and experience in data analysis varied. 
According to their self-ratings on a 5-point Likert scale (1: lowest extent, 5: greatest extent), they were experienced with the Pandas syntax used in ViseGPT (M = 3.28, SD = 0.57) and data wrangling (M = 3.17, SD = 0.92). 
Additionally, they had extensive experience in using LLM-powered natural language-based conversational interactions products (M = 4.33, SD = 0.59).
In summary, all participants met the participation requirements for the user evaluation.

\textbf{Baseline. }
We removed the extended views and associated derivative features in ViseGPT, designating the experimental group that subsequently used this system for debugging as Baseline. 
% We created a simplified version of ViseGPT by removing the extended views and associated derivative features.
In other words, the Baseline system is functionally equivalent to LLM-powered dialogue tools (e.g., ChatGPT) that participants commonly interact with.
% The baseline system functions similarly to standard LLM interfaces like ChatGPT.
To minimize variables and enhance the reliability of experimental conclusions, Baseline maintains consistent stylistic design and basic functionalities (e.g., conversation context preservation) with ViseGPT, while utilizing the same foundation model (i.e., llama-3-70b-instruct).
% By controlling for other factors, we maintained identical visual design and core functionalities, including conversation context preservation, and used the same foundation model (i.e., llama-3-70b-instruct) in both ViseGPT and Baseline.

\textbf{Tasks. }The evaluation included four tasks (A-D) sampled from distinct work contexts (refer to supplementary material for details).  
% In each task, we provided a piece of target data wrangling script description and an issue-embedded Pandas script. 
For each task, we presented participants with requirements of data wrangling scripts and a corresponding Pandas script containing functional errors.
Although these scripts were generated using LLM with the requirements as prompts, they were syntactically correct but deviated from the intended functionality.
% Each script could be compiled (i.e., no syntax errors), but had some mismatches with the target description. 
% Hypothetically speaking, this script was directly generated by feeding the description as a prompt to the LLM. 
% Participants were tasked with validating the script's correctness, identifying issues, and implementing appropriate modifications as quickly as possible with the assistance of either ViseGPT or Baseline. 
Participants were required to debug the scripts using either ViseGPT or Baseline by identifying and correcting discrepancies from requirements.
% All the issues embedded in the experiment-provided scripts were adapted from real-world cases collected during formative study 
All error cases in the experimental scripts were derived from real-world examples documented in our formative study (Section \ref{section:3}).

\subsection{Procedure} 
We opted for a modified balanced Latin square design \cite{DEAN20015090} to control for potential order effects of tool usage on task performance. 
% Specifically speaking, 
In our experiment design, each tool (ViseGPT and Baseline) appeared exactly 9 times in each task (A-D), collectively covering all possible tool-task combinations (Total: 6 groups $\times$ 4 tasks $\times$ 3 rounds = 72 trials, 36 trials per tool and 9 trials per tool for each task). 
This approach enabled statistically robust comparisons of tool performance across multiple tasks while controlling for learning biases and individual differences.
The complete experimental procedure lasted approximately 60 minutes.

% Prior to formal task initiation, 
Prior to experimental procedures, participants took part in a training session consisting of a system tutorial and a practice task, which took about 5 minutes.
% participants were briefed on the tools and completed a warm-up task.
% Throughout the process, they could ask for details. 
Subsequently, the participant worked on Task A-D sequentially, with each task allocated a 15-minute time limit. 
% In each task, participants stopped working either when they reached the time limit or when they believed debugging was complete (whichever came first).
Participants ended each task when they either reached the time limit or completed the debugging.
The entire task execution process was screen-captured for subsequent analysis. 
The evaluation ended with a semi-structured interview and a User Experience Questionnaire  (UEQ) \cite{10.1007/978-3-319-07668-3_37}. 
% During the interviews, participants were asked to describe their overall impressions of the tool's performance across two debugging stages and provide improvement suggestions. 
The interviews focused on gathering participants' overall impressions of tool performance across script debugging stages and eliciting improvement suggestions.
We adopted the UEQ 
% \cite{10.1007/978-3-319-07668-3_37}
to measure participants' perceived experience with both ViseGPT and Baseline, and compared the results using the official UEQ data analysis tool \cite{UEQ2024}. 
In total, the post-experiment interview and questionnaire took no longer than 15 minutes. 
The participation fee was \$10 per hour.

\subsection{Results}
We compare task correctness, time spent and participants' subjective ratings between Baseline and ViseGPT.
In addition, we analyze specific interaction behaviors of participants and report insights derived from the interview.

\subsubsection{Task correctness and spent time}
\label{section:6.3.1}
\autoref{table:result} compares participant performance between ViseGPT and Baseline conditions across the four experimental tasks (A-D). 
% We first focus on the ultimate success rates - defined as the proportion of cases where participants ultimately made correct modifications following debugging procedures. 
% Across all tasks, participants in the ViseGPT condition demonstrated higher success rates than those in the Baseline condition. 
The success rate, defined as the percentage of participants who successfully fixed the bug, was higher in the ViseGPT condition for Tasks A, B, and D. 
% Notably, success rate for ViseGPT in Task C were significantly lower compared to the other three tasks.
However, Task C showed comparable success rates between conditions (ViseGPT: 56\% vs. Baseline: 22\%).
Furthermore, as shown in \autoref{fig:time},
% shows that completion times were significantly shorter in the ViseGPT condition compared to Baseline 
ViseGPT significantly reduced completion times for Tasks A, B, and D (Mean difference > 120s), while no substantial difference was observed for Task C (ViseGPT: M = 625s vs. Baseline: M = 645s).

% We now analyze the factors underlying ViseGPT's relatively weaker performance in Task C. 
We analyze the factors underlying ViseGPT's lower performance in Task C.
% The experimental context of Task C required participants to calculate the global influence scores of \textbf{five specified colors} in images. 
The task required participants to calculate global influence scores for \textbf{five specified colors} in images.
The raw data used standard case-insensitive \#RRGGBB color codes, while the initial script lacked case normalization.
% but the task-provided script didn't normalize case variations. 
Consequently, the color sorting operation treated differently-cased representations (e.g., \#FF0000 and \#ff0000) as distinct colors, resulting in erroneous counts exceeding the five target colors. 
First, ViseGPT successfully generated test cases to verify that the \texttt{rank\_label} column (representing color influence rankings) should contain integers between 1 and 5, as specified in the prompt. 
Consequently, the system correctly flagged issues when \texttt{rank\_label} values exceeded 5. 
We observed that although all participants detected this issue, their subsequent analysis of the issues and the prompts they submitted for script regeneration impacted ViseGPT's performance. 
A minority of participants (3/9) promptly identified the range issues arising from duplicate color code counts. 
By directly instructing ViseGPT to normalize color code casing, they resolved the issue. 
However, the majority of participants (6/9) opted to describe the situation directly to ViseGPT (e.g.,\textit{``The rank\_label column contained invalid values exceeding the upper bound of 5.''}) without addressing the underlying cause of color duplication. 
This led ViseGPT to implement suboptimal solutions, such as using \texttt{pd.head()} to limit output to five rows, instead of standardizing the color format.
This resulted in extended debugging cycles and reduced success rates.
% The system's superficial interpretation of participant intent led it to erroneously employ \pythoninline{pd.head()} to simply retrieve the first five rows of data, rather than implementing the required format standardization. 
% This flawed approach trapped participants in inefficient iteration cycles, consequently increasing time expenditure and reducing debugging success rates.

\subsubsection{Subjective ratings} 
After UEQ investigation, we used Shapiro-Wilk tests to test normality (all $p>0.05$), and used Student's paired t-tests ($\alpha=0.05$).
The results demonstrated that ViseGPT was significantly better ($p<0.05$) in all six dimensions compared with Baseline (\autoref{fig:UEQ}). 
% All six dimensions (attractiveness, perspicuity, efficiency, dependability, stimulation and novelty) showed significant differences . 
% At a alpha-level of 0.01, all dimensions except perspicuity still maintained significant differences.

\subsubsection{Specific interaction behaviors}
\label{section:6.3.3}
We recorded three interaction metrics during the study (\autoref{table:result}): (1) the number of queries sent, (2) the number of queries with test reports attached in ViseGPT, and (3) counts of clicking rectangles in the tailored Gantt chart. 

Our analysis revealed that participants preferred attaching test reports when submitting prompts in ViseGPT across all four tasks, with this behavior accounting for over 60\% of total query submissions when using ViseGPT. 
The situation demonstrates that participants in the ViseGPT workflow showed a preference for uploading test results as supplementary material when submitting prompts. 
We infer that this extended functionality reduces users' manual issue-summarizing efforts, while its interactive usability could potentially enhance confidence in script modifications. 

We also observed that in all the tasks except Task D, participants sent more prompts in the Baseline condition than ViseGPT. 
With ViseGPT's intuitive visual debugging interface and advanced interactions, participants were typically able to obtain correct scripts within fewer iterations. 
For Task D, we noticed that two participants (P17 and P18) using the Baseline requested early termination after sending just one prompt, but their final scripts still contained issues.
% , which may have affected the comparison results. 
During interviews, we specifically inquired about this situation. 
Both stated that they ended the task because the Baseline system indicated the provided script had no issues, and they detected no problems during manual inspection.

\subsubsection{Overall impressions around two stages. }
\label{section:6.3.4}
% After the tasks ended, we interviewed the participants about their overall impressions.
Upon completion of all tasks, we conducted semi-structured interviews to evaluate participants' experiences.
The interviews specifically focused on participants' debugging processes across two critical stages: (1) script verification and (2) code modification implementation.
% The interviews focused on the two key stages of debugging: (1) verifying script correctness and (2) implementing script modifications. 

\textbf{About verifying script correctness and analyzing issues. }
The majority of participants (16/18) reported that ViseGPT's visualization provided more intuitive debugging assistance compared to Baseline's generic responses, facilitating faster issue identification. 
\textit{``I can directly click the red rectangles (in the tailored Gantt chart) to view failed test cases rather than reviewing the entire script from scratch.''}  (P12)
Additionally, participants (12/18) appreciated the automatically generated test cases for their logical categorization, which enhanced debugging precision. 
Notably, some of participants (3/18) raised concerns about potential undetected issues persisting in scripts despite full test case validation. 
\textit{``Due to inherent limitations in constrained data description methods, the test cases may lack rigor and fail to achieve full coverage.''} (P3)
And participants (P2 and P9) noted that compared to Baseline's textual explanations, ViseGPT offered less flexibility in elucidating script logic. 

\textbf{About modifying the scripts iteratively. }
Participants (17/18) reported that ViseGPT enabled context-aware rapid corrections, significantly boosting their confidence in script modifications. 
The auto-generated test cases (7/18) and test report attachment feature (14/18) reduced descriptive burden during debugging. 
However, some users (6/18) noted that Baseline's free-form questioning allowed direct explanation of modification logic, whereas ViseGPT's revision process was harder to track. 
\textit{``Although I didn't comprehend the source of issues, ViseGPT successfully corrected based solely on my description of the situation - but the lack of explanatory feedback left me somewhat confused about the resolution process. ''} (P14)

\subsubsection{Suggestions from participants. }
\label{section:6.3.5}
To enhance ViseGPT's practical utility, we collected optimization suggestions from participants during user studies.

\textbf{Expand test case categories and severity classification.}
Beyond the 8 predefined test case categories of ViseGPT, participants recommended expanding more categories (6/18) and even support user-defined test case category (3/18) to improve test coverage and flexibility. 
Moreover, some participants (2/18) expressed that the current color-coding scheme - using red for issues and green for passed tests - appears overly simplistic for complex debugging scenarios.
\textit{``ViseGPT could adopt modern compiler conventions by classifying less severe issues as warnings and marking them in yellow, thereby developing a three-tier status indicator system.'' }(P17)
With severity classification, users can prioritize debugging efforts, focusing first on critical issues while deprioritizing minor warnings.

\textbf{Implement data causality analysis to enhance tracing.} 
Participants (P5, P10 and P13) pointed out that ViseGPT's insufficient presentation of dataflow causality made it difficult to trace the transformation process of erroneous data during debugging. 
For example, in Task D, the erroneous value 20.5 in the \texttt{purchase\_amount} column was originally positioned mid-column. 
However, after executing \texttt{df.sort\_values(by='purchase\_amount'}, the value relocated to the column's final row. 
\textit{``Tracing the data transformation process consumed significant time - while I could identify the erroneous output, determining its origin proved still challenging.'' }(P10)
Therefore, in addition to the existing single-step data display, providing visualization of erroneous data's transformation path across steps would improve debugging efficiency.
\section{Discussion}
This paper introduces ViseGPT, a novel tool for enhancing the efficiency of users in debugging LLM-generated data wrangling scripts, aiming to better align the scripts with user prompts. 
User evaluation revealed that using ViseGPT makes issue validation in scripts more intuitive and boosts confidence in correctness after modifications (Section \ref{section:6.3.4}). 
In this section, we discuss the implications of the system design (Section \ref{section:6.1}) and outline the limitations of the research as well as future work directions (Section \ref{section:6.2}).

\subsection{Design Implications}
\label{section:6.1}
Throughout the design and development of ViseGPT, we have gained valuable implications  beyond the system itself.

\textbf{Implementing a ``testing-as-conversation'' interaction model to refine debugging workflow. }
While traditional testing remains a post-development phase \cite{5315981}, ViseGPT compresses the entire data wrangling script development process into a simplified workflow, using a ``testing-as-conversation'' paradigm \cite{Hubley02102017}. 
This design approach adheres to M. Fowler's ``Continuous Integration'' principle \cite{fowler2006continuous}, which is one of the core practices in Agile development \cite{shore2021art}. 
First, developers can immediately transform failing cases into debugging clues through interactive workflows. 
\textit{``I believe that even users with limited (data wrangling script) programming expertise can derive novel insights into script debugging with ViseGPT.''} (P11) 
Based on this perspective, ViseGPT adapts to users with varying levels of expertise in data wrangling script development.

\textbf{Establishing a framework of constraints enhances clarity and boosts users' confidence. }
ViseGPT systematically extracts and validates constraints from natural language prompts.
By making expectations explicit, such a framework reduces ambiguity in both the user intent and the behavior of generated scripts. 
This approach aligns with the ``Specification as Code'' paradigm  \cite{205185,10.1145/195274.195297} from formal methods. 
Interacting with a well-structured testing framework can enhance users’ confidence in the final results \cite{liu2011framework}. 
\textit{``The end-to-end process—from sending a prompt, to generating results, to validating results with the original prompt—serves as a confidence bridge between me and ViseGPT.''} (P16)
The design tries to expand the trust boundaries of Human-LLM collaboration.

\textbf{Enhancing debug reasoning through visualization. }
Observing how users debug LLM-generated scripts (Section \ref{section:6.3.3}), we found that visualization in ViseGPT serves not just as an infomation presentation medium, but as a central mechanism for supporting logical reasoning and identifying issues. 
Unlike conventional debugging tools that provide textual descriptions of results and counts of bugs \cite{10.1145/251880.251926}, the tailored Gantt chart summarizes the execution process, enabling users to quickly identify critical steps and spot issues. 
\textit{``During debugging, observing visual elements is more engaging and effective than focusing on code. ''} (P7)
This opinion serves as validation for attraction of our visualization design. 
However, we should remain mindful of its inherent limitations, possibly introducing new issues rather than resolving them (Section \ref{section:6.3.1}).
% \textbf{Leveraging visualization techniques to construct cognitive interfaces with hybrid mental models. }
% The hybrid mental model represents a cognitive framework where users simultaneously engage in intuitive experiential reasoning (e.g., association and pattern matching) and symbolic reasoning (e.g., rules and structured knowledge) when interacting with AI systems \cite{du2018voice, Fernando04072022}. 
% Specifically, in ViseGPT, the tailored Gantt chart's vertical flow mapping aligns with human intuition of "data wrangling pipelines", facilitating precise issue localization. 
% And color and shape encodings test cases and results as visual signals, which can activate users' innate pattern recognition capabilities \cite{lindberg2020developing}. 
% It may be attributed as one key factor behind ViseGPT's superior debugging performance demonstrated in Section \ref{section:6.3.1}, also validating the continued relevance of Don Norman's design principles for cognitive artifacts \cite{norman1992design} in the AI era.

\subsection{Aligned Visualizations for AI Explainability}
Building upon our visualization design for script debugging (Section \ref{section:5}), we argue that the aligned-to-output-line paradigm can bridge the ``gulf of evaluation'' in Human-LLM collaboration across domains \cite{SWELLER1988257}. 
Norman's gulfs of execution and evaluation \cite{norman1986cognitive} highlight the cognitive gap between system outputs and user interpretation, which is also a challenge acutely present in LLM interactions where users struggle to trace how outputs relate to their intent.
ViseGPT's Gantt chart displays that structural alignment between output components (code lines) and their visual explanations (test results) enhances interpretability. 
The key point of this approach lies in its ability to externalize the implicit relationships between input prompts, intermediate reasoning, and final outputs. 
Our participant feedback (Section \ref{section:6.3.4}) supports that such spatial organization aids pattern recognition and issue localization, which is transferable to scenarios requiring stepwise validation of LLM outputs.

In conclusion, the ``aligned-to-output-line'' strategy offers a scaffold for cross-domain explainability standards.
By adopting this paradigm, tools could mitigate the "black-box" effect of LLMs, transforming opaque reasoning chains into inspectable, interactive artifacts \cite{doshivelez2017rigorousscienceinterpretablemachine}.
We envision extensions of this work adapting the alignment principle to multimodal outputs (e.g., image-text pairs) or collaborative settings where visual explanations serve as shared referents for team debugging—ultimately advancing the goal of human-AI interaction transparency.

\subsection{Limitations and Future Work}
\label{section:7.3}
This section examines three key limitations of the ViseGPT system in data wrangling script debugging and proposes potential directions for future work.

\textbf{Step Granularity Trade-offs.} The step-based segmentation in ViseGPT follows an ``aligned-to-line'' design (Section \ref{section:5.2}). However, this granularity may introduce scalability limitations: long scripts suffer from reduced readability due to vertical sprawl, while complex single-step operations (e.g., multi-column transformations) may be hard to understand in a step-level view. Future work can investigate a framework with hierarchical abstractions, balancing coarser block-level summaries for scalability with finer sub-step breakdowns for intricate operations.

\textbf{Task Representativeness and Complexity.} 
In the evaluation, the four tasks (Section \ref{section:6.1}) were selected from formative study to ensure practical relevance and they focus on common violations (e.g., formats, ranges) that directly manifest in output data, as such issues represent high-frequency pain points identified.
However, we acknowledge that real-world data wrangling scripts often involve more complex situations.
The current evaluation demonstrates ViseGPT's effectiveness in handling common debugging scenarios, but future work should expand to include more diverse and sophisticated issue types to further validate the versatility.

\textbf{Test Case Coverage and Flexibility. }
While ViseGPT's automated test case generation improves debugging accuracy efficiency, its current framework is limited to predefined constraint categories as a concern expressed by the participants in Section \ref{section:6.3.4}. 
This may not fully capture nuanced or domain-specific requirements, leading to potential gaps in test coverage. 
Future work could expand the system's flexibility by allowing users to define custom test case categories or integrate dynamic learning \cite{zolitschka2020novel} from iterative feedback, thereby enhancing adaptability to diverse scenarios.

\textbf{Language and Framework Generalization. }
Currently, ViseGPT primarily supports Python-based data wrangling scripts, limiting its applicability to users working with other programming languages (e.g., R, SQL, or Julia). 
Future research should explore extending the system’s requirement analysis and test generation capabilities to additional languages, as well as integrating with domain-specific frameworks.
% (e.g., PySpark \cite{Spark2025} for big data processing). 
This expansion would broaden ViseGPT’s utility across different programming ecosystems.

\textbf{Debugging Transparency and Workflow Integration. }
The tool's visualization effectively highlights test failures but lacks detailed tracing of data causality across script steps (Section \ref{section:6.3.5}), which can complicate root cause analysis. 
Additionally, its standalone nature may disrupt existing workflows, such as integration with IDEs or version control systems. 
Future work could enhance debugging transparency with interactive dataflow graphs and improve usability through seamless integration with popular development environments (e.g., Jupyter Notebook, VS Code) and collaborative features for team-based debugging.
\label{section:6.2}
\section{Conclusion}
This paper presents ViseGPT, a tool that addresses the challenge of debugging LLM-generated data wrangling scripts by achieving better alignment between natural language prompts and results. 
Through automatically extracting constraints from user instructions and generating systematic test cases, ViseGPT enables users to efficiently verify whether generated scripts align with their requirements without requiring deep programming expertise. 
The system's constraint-based validation approach, combined with its intuitive Gantt-chart visualization and interactive debugging features, provides users with actionable insights to identify and resolve script issues. 
Our user evaluation (N=18) demonstrates that ViseGPT improves the efficiency and accuracy of debugging LLM-generated data wrangling scripts, making complex tasks more accessible. 
This work advances the field of human-AI collaboration by offering a practical solution that enhances both the trustworthiness and iterability of AI-generated code, ultimately empowering users to harness the full potential of LLMs with greater confidence.

\begin{acks}
The authors would like to thank the anonymous reviewers for their constructive comments. This work was supported in part by Zhejiang Provincial Natural Science Foundation of China under Grant No. LD25F020003, in part by NSFC (U22A2032, 62402421), and in part by Ningbo Yongjiang Talent Programme (2024A-399-G).
\end{acks}
%% the bibliography file.`
\bibliographystyle{ACM-Reference-Format}
\bibliography{references}

\end{document}